\documentclass[conference]{IEEEtran}
\usepackage{cite}
\usepackage{physics}
\usepackage{amsmath,amssymb,amsfonts,hyperref}
\usepackage{algorithm}
\usepackage{algpseudocode}
\usepackage{graphicx}
\usepackage{textcomp}
\usepackage[usenames,dvipsnames]{xcolor}
\usepackage[noabbrev,capitalise]{cleveref}
\hypersetup{colorlinks,linkcolor=black,citecolor=red,urlcolor=blue}

\def\BibTeX{{\rm B\kern-.05em{\sc i\kern-.025em b}\kern-.08em
    T\kern-.1667em\lower.7ex\hbox{E}\kern-.125emX}}
\begin{document}

\title{Low-Depth Flag-Style Syndrome Extraction for Small Quantum Error-Correction Codes}
\author{\IEEEauthorblockN{Dhruv Bhatnagar*}
\IEEEauthorblockA{\textit{Quantum \& Computer Engineering Department} \\
\textit{Delft University of Technology}\\
2628 CD Delft, The Netherlands \\
\href{mailto:d.bhatnagar@student.tudelft.nl}{d.bhatnagar@student.tudelft.nl}}
\and
\IEEEauthorblockN{Matthew Steinberg*}
\IEEEauthorblockA{\textit{Quantum \& Computer Engineering Department} \\
\textit{Delft University of Technology}\\
2628 CD Delft, The Netherlands \\
\href{mailto:matt.steinberg3@gmail.com}{matt.steinberg3@gmail.com}}
\and
\IEEEauthorblockN{David Elkouss}
\IEEEauthorblockA{\textit{Networked Quantum Devices Unit} \\
\textit{OIST Graduate University}\\
Okinawa, Japan \\
\href{mailto:david.elkouss@oist.jp}{david.elkouss@oist.jp}}
\and
\IEEEauthorblockN{Carmen G. Almudever}
\IEEEauthorblockA{\textit{Computer Engineering Department} \\
\textit{Technical University of Valencia}\\
Valencia, Spain \\
\href{mailto:cargara2@disca.upv.es}{cargara2@disca.upv.es}}
\and
\IEEEauthorblockN{Sebastian Feld}
\IEEEauthorblockA{\textit{Quantum \& Computer Engineering Department} \\
\textit{Delft University of Technology}\\
2628 CD Delft, The Netherlands \\
\href{mailto:s.feld@tudelft.nl}{s.feld@tudelft.nl}}}

\maketitle
\def\thefootnote{*}\footnotetext{These authors contributed equally to this work. The corresponding author for this work is Matthew Steinberg.}\def\thefootnote{\arabic{footnote}}

\begin{abstract}
Flag-style fault-tolerance has become a linchpin in the realization of small fault-tolerant quantum-error correction experiments. The flag protocol's utility hinges on low qubit overhead, which is typically much smaller than in other approaches. However, as in most fault-tolerance protocols, the advantages of flag-style error correction come with a tradeoff: fault tolerance can be guaranteed, but such protocols involve high-depth circuits, due to the need for repeated stabilizer measurements. Here, we demonstrate that a dynamic choice of stabilizer measurements, based on past syndromes, and the utilization of elements from the full stabilizer group, leads to flag protocols with lower-depth syndrome-extraction circuits for the [[5,1,3]] code, as well as for the Steane code when compared to previously-established methods in flag fault tolerance. We methodically prove that our new protocols yield fault-tolerant lookup tables, and demonstrate them with a pseudothreshold simulation, showcasing large improvements for all protocols when compared to previously-established methods. This work opens the dialogue on exploiting the properties of the full stabilizer group for reducing circuit overhead in fault-tolerant quantum-error correction.
\end{abstract}



\section{Introduction}

Flag-style fault-tolerant quantum-error correction gives rise to syndrome extraction with fewer ancilla qubits than other typical fault-tolerance protocols (FT) in quantum-error correction \cite{arbitrarychamberland,terhalReview,cross2009comparativeFT}. First developed in \cite{chaoreichardt18,chao18FTcomputation,Yoder17twist}, the protocol allows for higher-weight errors to propagate through an extra ancilla qubit which is known as the \textit{flag qubit}. It has been shown that this additional ancilla qubit permits the efficient determination of the most-likely corresponding low-weight errors, as well as correction of higher-weight errors in subsequent error-correction rounds. 

Having been extended to many different stabilizer codes within the fault-tolerance domain, as well as for fault-tolerant state preparation and quantum computation, among others\cite{PRXQchaoANYSTAB,Yoder17twist,gd1,gd2,gd3,gd4,gd5,gd6,gd7,gd8,gd9,gd10,gd11,gd12,gd13,gd14,gd15}, the flag protocol has enabled the realization of several experiments demonstrating fault-tolerant quantum-error correction techniques using current \textit{noisy intermediate-scale quantum} (NISQ) hardware \cite{experiment1,experiment2,experiment3,experiment4}. The flag protocol has in particular found great applicability in stabilizer codes, where one usually requires a step of classical processing in order to ascertain which stabilizer generator to measure next in an iterative sequence. This technique is known as \textit{adaptive syndrome extraction} \cite{arbitrarychamberland,chaoreichardt18}. In most cases, central to the protocol is the assumption that reset and measurement operations for ancilla qubits can be completed relatively quickly, although adaptations of the original flag protocol can address this issue with static syndrome-extraction sequences for architectures with slower qubit-readout and reset times \cite{gd5}. In this paper, we focus on the regime of fast measurement and reset operations; we reserve the consideration of our new protocols for the slow measurement and reset regime in future work.

In most fault-tolerance protocols currently known, stabilizer measurements are repeated several times, in order to ensure accuracy of the syndrome- and flag-qubit measurements; this constitutes a large burden in terms of additional gate overhead and depth costs associated with executing a quantum algorithm on a quantum computer \cite{cross2009comparativeFT}. However, in a recent paper \cite{delfosse2020short}, it was shown that the gate overhead for Shor-style syndrome-extraction circuit sequences could be greatly reduced for a large number of stabilizer-code classes, including the well-known CSS codes. For the Steane code in particular, one usually requires up to $24$ stabilizer measurements in order to guarantee fault tolerance. By analyzing the possible errors contingent upon error propagation, it was systematically shown that the length of the Shor protocol could be reduced to only $7$ stabilizer measurements, if one considers not only the six typical stabilizer generators, but additionally higher-weight, mixed-qubit support, and mixed-Pauli stabilizer elements which reside in the full stabilizer group of $2^{(n-k)}$ elements.

In this work, we adapt the analysis proposed in \cite{delfosse2020short} to the context of flag fault-tolerant quantum-error correction. As flag fault-tolerant quantum-error correction suffers from the same gate-overhead issues as other known fault-tolerance protocols \cite{cross2009comparativeFT}, we show that the syndrome-extraction sequences for the $[[5,1,3]]$ code can be reduced in the case of propagated-error detection. Additionally, we find that the $[[7,1,3]]$ Steane code's flag syndrome-extraction circuit can also be reduced for the case of non-trivial syndrome- and flag-qubit measurement outcomes. We show several example protocols which are required to ensure a fault-tolerant lookup table (LUT), while providing a decrease in the range of $25-50\%$ fewer two-qubit gates in a particular subround when compared to measuring all of the stabilizers in the same subround from the protocol in \cite{chaoreichardt18}. We systematically test these new protocols against the state-of-the-art procedure from \cite{chaoreichardt18}, and find pseudothreshold improvements in the range of $3.64\% - 11.16\%$ as a result. We believe that further refinement of our proposal is possible, as well as generalizations to other codes. 

The remainder of this paper is structured as follows: \cref{section:background} summarizes the original flag protocol and introduces relevant terminology, as presented in \cite{chaoreichardt18}; \cref{section:shortFlagprotocols} presents the structure of our new flag fault-tolerance protocols; \cref{section:results} shows the results and analysis of our pseudothreshold simulation, performed using Qiskit \cite{Qiskit}; finally, we provide concluding comments in \cref{section:conclude}.

\section{Background} \label{section:background}

\begin{figure}
\centering
\includegraphics[width=7cm]{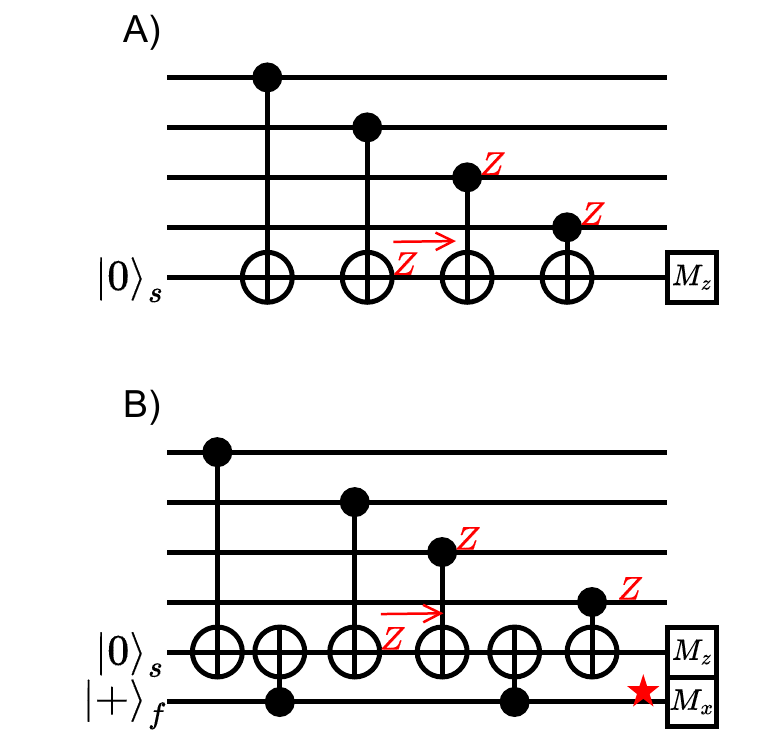}
\caption{Two syndrome-extraction circuits are shown in A) and B) for measuring the syndromes of the stabilizer $ZZZZ$. A) is not fault tolerant, as the propagation of a $Z$-error can perniciously lead to a logical error which is undetectable; B) however, is fault tolerant, as the propagated errors are detected by the flag-qubit measurement, which is carried out in the $X$-basis. The red star marks the detection of the propagated error via $X$-basis measurement. Syndrome-qubit measurement is realized in the $Z$-basis for the protocols under consideration.}
\label{fig:flagprotocol}
\end{figure}

Two simple syndrome-extraction circuits for one of the stabilizers of the Steane code is shown in \cref{fig:flagprotocol}. The core of the flag technique involves appending an additional qubit to the syndrome qubit with the goal of tracking propagated errors. As shown in \cref{fig:flagprotocol}A), an imperfectly executed CX gate can spawn an extra $Z$ error. This $Z$ error undergoes unitary evolution and evolves into a $ZZ$ error after each of the following two CX gates. Upon measuring the syndrome qubit in the $Z$ basis, the propagated errors will not be detected. As such, a possible solution is to add one extra ancilla qubit (in the case of distance-3 codes), prepare it in the $\ket{+}$ state, and entangle it with the original syndrome qubit (prepared in the $\ket{0}$ state); this is shown in \cref{fig:flagprotocol}B), and is known as a \textit{flag qubit}. In this setting, the errors can propagate throughout the circuit, but due to the properties of the entangled \textit{flag qubit}, a measurement in the $X$-basis reveals that an error has propagated (marked with a red star in \cref{fig:flagprotocol}). 

\begin{figure}
\centering
\includegraphics[width=8cm]{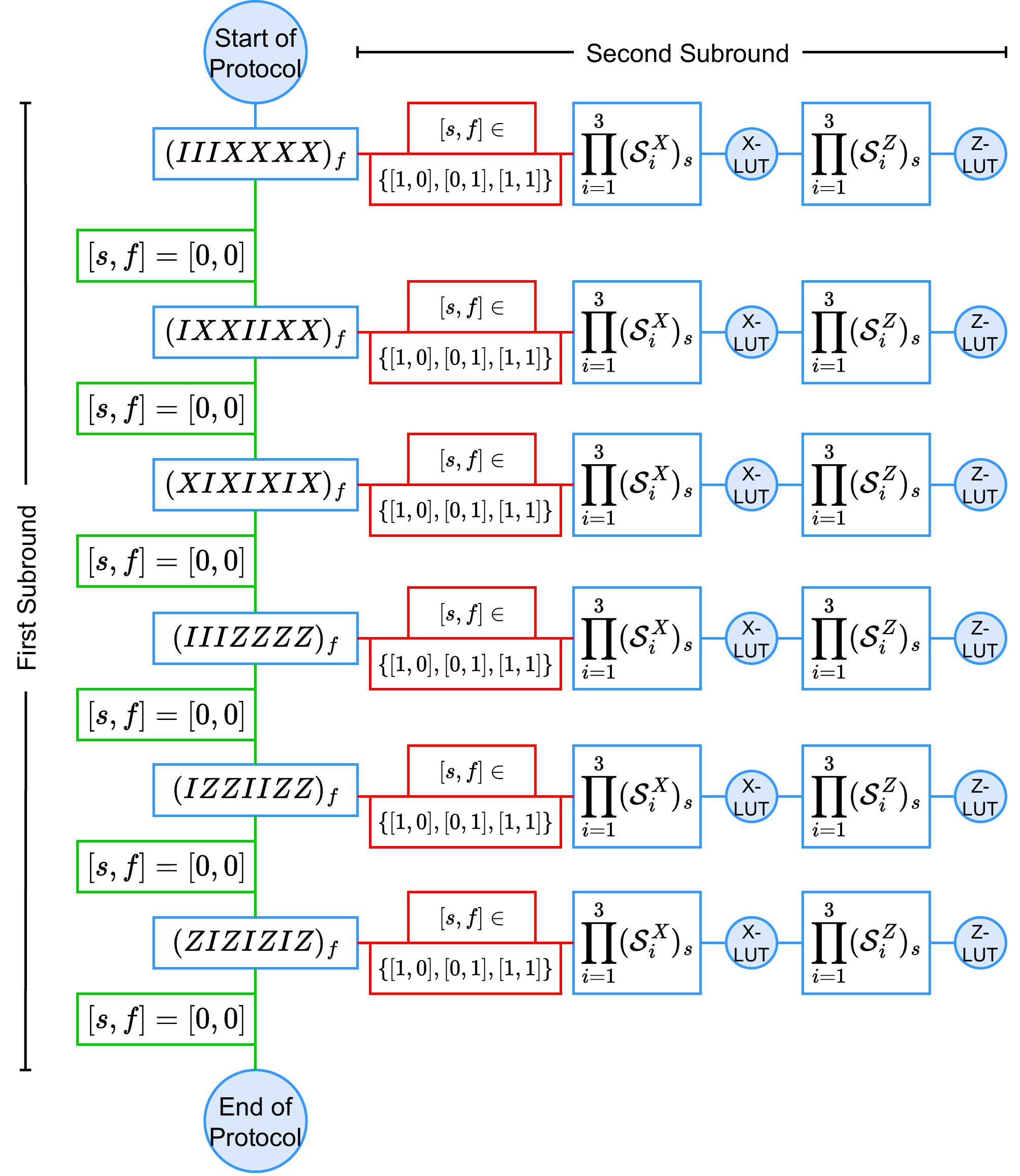}
\caption{A diagram depicting the original flag fault-tolerance protocol from \cite{chaoreichardt18} for the Steane code. Here, the starting and endpoints of the protocol are denoted with blue circles on the top and bottom of the figure. Green lines emphasize trivial syndrome/flag measurements in the first subround, and subsequent flagged stabilizer measurements; the notation $(\cdot)_{f}$ denotes a flagged stabilizer, while $(\cdot)_{s}$ denotes unflagged stabilizers. Red lines highlight non-trivial syndrome/flag measurements in the first subround, which lead to unflagged stabilizer measurements in the second subround. Concordant with typical decoding schemes for CSS codes, unflagged $X$-type stabilizer measurements and corrections are applied first via a lookup table (LUT), with $Z$-type measurements and corrections afterward. After performing an LUT correction, the protocol effectively ends; alternatively, if the first subround terminates with all syndrome pairs $[s,f]$ trivial, then the protocol ends, as well.}
\label{fig:steane_decisiontree_CR18}
\end{figure}

\begin{figure}
\centering
\includegraphics[width=6.5cm]{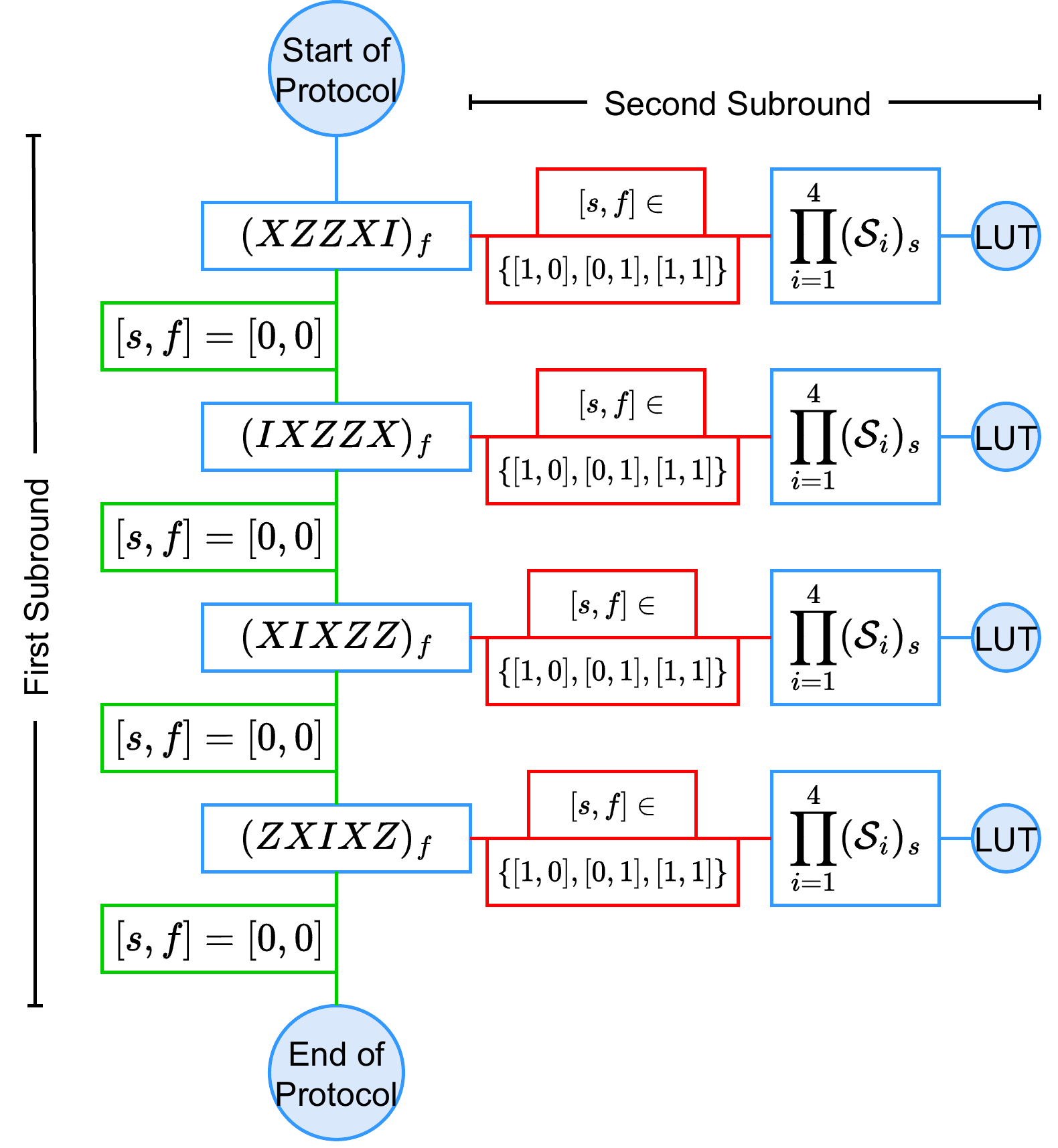}
\caption{A diagram depicting the original flag fault-tolerance protocol from \cite{chaoreichardt18} for the $[[5,1,3]]$ code. Here, most of the protocol is similar or the same as in \cref{fig:steane_decisiontree_CR18}; however no separation between $X$- and $Z$-type stabilizers is given due to the structure of the $[[5,1,3]]$ code. As such, decoding in the second subround is achieved by accounting for both Pauli error types in the same LUT.}
\label{fig:[513]_decisiontree_CR18}
\end{figure}

However, measuring only one stabilizer generator is not sufficient in order to completely identify and diagnose an error. In \cite{chaoreichardt18}, two \textit{subrounds} of six stabilizer measurements are performed; the first subround is executed with an additional flag ancilla, and the second, carried out only in the event of a non-trivial syndrome- or flag-qubit measurement (i.e. if syndrome- and flag-qubit measurement outcomes are $[s,f] \in \{[1,0],[0,1],[1,1]\}$, where $[s,f]$ refers to the syndrome-and flag-qubit measurements), is implemented with only syndrome qubits. \cref{fig:steane_decisiontree_CR18,fig:[513]_decisiontree_CR18} display the original protocols from \cite{chaoreichardt18}. In this work, we construct three new fault-tolerant protocols with the same general structure.

The original protocol from \cite{chaoreichardt18} for the Steane and $[[5,1,3]]$ codes proceeds as follows. In the first subround, depicted on the left-hand column of \cref{fig:[513]_decisiontree_CR18}, the stabilizer generators are measured with the extra flag ancilla qubit. If the $Z$- and $X$-measurement outcomes are trivial (i.e. $[s,f] = [0,0]$), then the next generator is successively measured, until all $(n-k)$ generators which define the code are measured. If the measurement outcomes $[s,f]$ are non-trivial, then the first subround ends, and a second subround begins; here, all generators are measured once again, but without flag ancilla qubits. Once all generators are measured and syndromes are stored classically, decoding via an LUT commences. In the case of CSS codes, decoding is performed separately for $X$- and $Z$-type stabilizer generators, as shown in \cref{fig:steane_decisiontree_CR18} and labeled as "X-/Z-LUT", to denote corrections applied after measuring $X$-/$Z$-type stabilizers. The original flag protocol is displayed for two example codes, the $[[5,1,3]]$ and the Steane codes, in \cref{fig:[513]_decisiontree_CR18,fig:steane_decisiontree_CR18}.

As is evidenced from \cref{fig:[513]_decisiontree_CR18,fig:steane_decisiontree_CR18}, the original flag protocol utilizes only the $(n-k)$ stabilizer generators used to define a stabilizer code. In the following section, we detail our syndrome-extraction circuit reductions by employing particular elements of the stabilizer group which are not generators, i.e. other stabilizer operators from the stabilizer group of size $2^{(n-k)}$ elements.

\section{Shortened Flag Protocols using Elements of the Full Stabilizer Group} \label{section:shortFlagprotocols}

Although the original protocols from \cite{chaoreichardt18} paved the way for the first experiments in fault tolerance, such protocols are not optimal in the numbers of minimal and maximal two-qubit gate counts, as complete syndrome-extraction circuits require between $4$ and $8$ stabilizer measurements for the $[[5,1,3]]$ code, and between $6$ and $12$ syndrome measurements for the Steane code. This results in circuit gate counts between $24$ and $40$ two-qubit gates for the $[[5,1,3]]$ code, and between $36$ and $60$ for the Steane code.

The stabilizer group contains a total of $2^{(n-k)}$ stabilizer operators, of which a particular subset can be utilized in order to create a fault-tolerant syndrome-extraction circuit. Using this approach, reductions in the total number of two-qubit gate can be devised. We present here three examples of possible reductions. These shorter sequences apply only to each of the subrounds proposed in \cite{chaoreichardt18}, and we address separately the cases of a non-trivial syndrome-qubit measurement outcomes versus a non-trivial flag-qubit measurement outcomes.

\subsection{Fault-Tolerance Reductions for the [[5,1,3]] Code}

\begin{figure}
\centering
\includegraphics[width=\columnwidth]{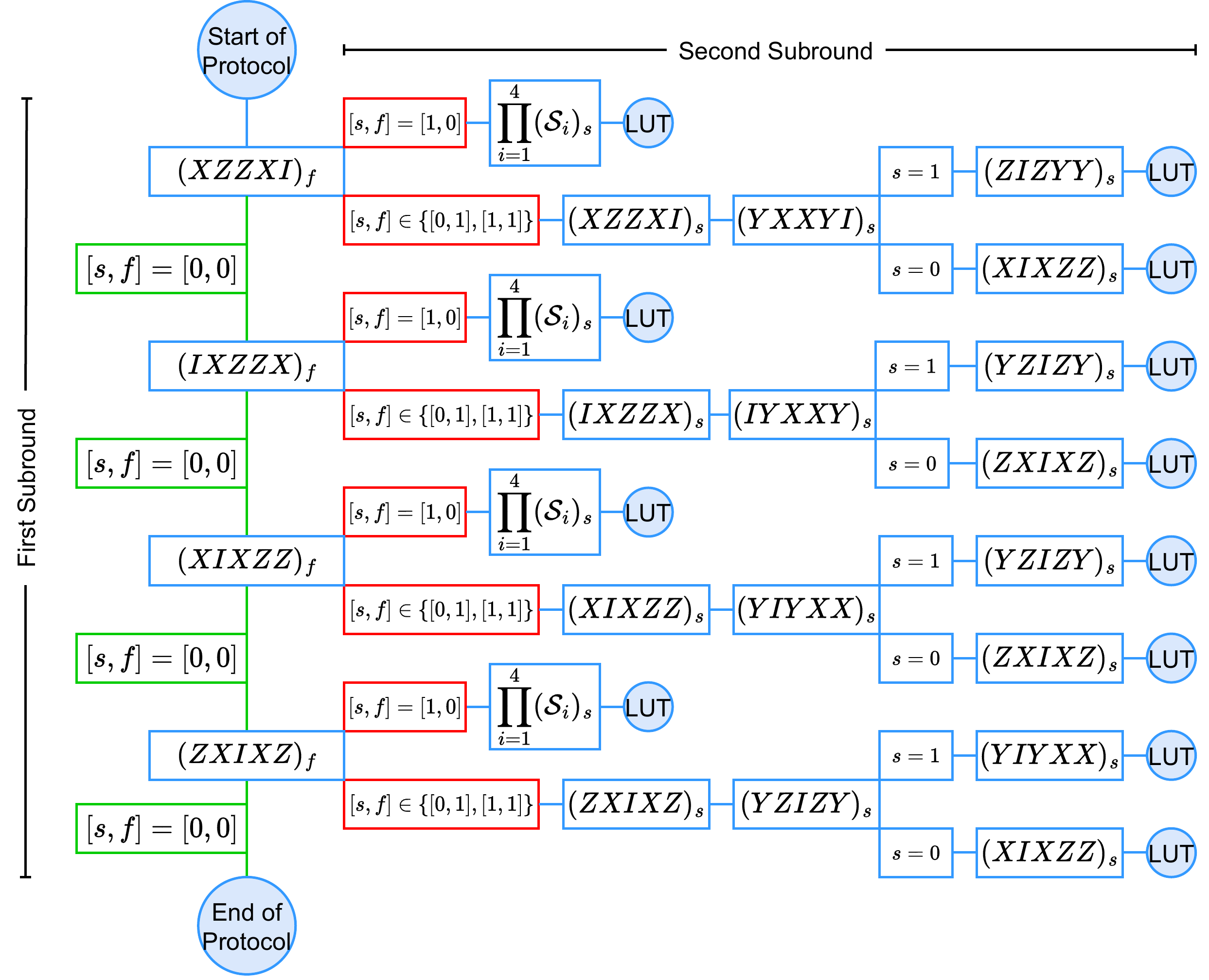}
\caption{The new $[[5,1,3]]$ code protocol. Note that our protocol utilizes $25\%$ less gates than the original method proposed in \cite{chaoreichardt18} when propagated errors are detected. When no errors are detected, the protocol proceeds in exactly the same fashion as in the original protocol.}
\label{fig:[513]_decision_new}
\end{figure}

Let us first consider the syndrome-extraction circuit for the $[[5,1,3]]$ code; a diagram for the protocol is shown in \cref{fig:[513]_decision_new}. In the protocol proposed in \cite{chaoreichardt18}, the first subround begins by measuring the four weight-$4$ ($w=4$) stabilizer generators $\{XZZXI, IXZZX, XIXZZ, ZXIXZ\}$ with flag-qubit ancillas; if a non-trivial syndrome is detected for the flagged generator $XZZXI$, we proceed by measuring the four same generators without the flag in the second subround. However, if there is a non-trivial flag-qubit measurement outcome, then we measure again the same stabilizer in the second subround without a flag, and then proceed to measure the stabilizer $YXXYI$; the measurement outcome of this stabilizer then leads to a third subround, in which we measure $ZIZYY$ in the case of a trivial syndrome ($s = 0$), and $XIXZZ$ for the case when $s = 1$. Such a variation can be repeated for all of the $[[5,1,3]]$ code's generators, in which the third stabilizer in the subround measured always shares the same qubit support as the flag-triggering stabilizer from the first subround (i.e. a stabilizer generator $XZZXI$ measured first gives rise to measuring $YXXYI$ third). 

It can be shown that the resulting LUT is fault tolerant, as all resulting syndromes are unique and non-trivially map to lowest-weight Pauli corrections. In particular, the second subround in \cite{chaoreichardt18} requires that all four generators must be again measured without flags. The resulting four-bit syndromes are unique and non-trivial. However, as an example, there are only seven possible inequivalent propagated errors indicated by the non-trivial flag for the first stabilizer; these errors are: $IIZXI, IXZXI, IYZXI, IZZXI, IIIXI, IIXXI,$ and $IIYXI$. As the classical Shannon entropy provides a lower bound on the number of bits needed to distinguish Pauli errors \cite{nielsen1-IT}, one can show that three bits (and hence three stabilizer measurements) are needed to distinguish these remaining errors. To ensure fault tolerance, inequivalent errors are mapped to distinct, non-trivial syndrome patterns, so that the trivial syndrome corresponds to a measurement error. We observe that these Pauli checks have different operators only on qubits two and three (from left to right), so the stabilizers chosen to distinguish them must have support on at least one of these qubits. The support of the stabilizer on the remaining qubits does not affect the syndrome information used to differentiate them. Thus, in contrast to the approach from \cite{chaoreichardt18}, we design our second-round unflagged stabilizer checks by first measuring a stabilizer with support on both qubits two and three, and then following with a corresponding stabilizer from outside the standard generator set, since the Pauli support on qubits two and three can be different. The last stabilizer is chosen such that support is contained only on qubit two or three. 

As shown in \cref{fig:[513]_decision_new}, our protocol reduces the gate count by $25\%$ for the second subround, in the case of propagated errors. For the case of no errors, the protocol proceeds exactly as the original, by measuring all four stabilizer generators. As the number of faulty positions in the circuit are decreased, we expect that an increase in the pseudothreshold should be discernible as well, and indeed this is the case, as discussed in \cref{section:results}.

\subsection{Fault-Tolerance Reductions for the Steane Code}

More invasive changes can be made in the case of the $[[7,1,3]]$ Steane code. For example, when measuring the first stabilizer generator, $IIIXXXX$, with a flag gadget, the resulting inequivalent correlated errors, indicated by a non-trivial flag measurement, belong to the set $IIIIIXX, IIIIXXX, IIIIYXX, IIIIZXX, \\ IIIIIIX, IIIIIYX, IIIIIZX$. Due to the weight $w$ of the stabilizer ($w = 4$ for the Steane code generators), as well as the flag qubit, we again count seven possibilities, and construct the subsequent stabilizer measurements such that these can be distinguished with three syndrome bits. Fault tolerance again requires that the syndromes corresponding to these errors be non-trivial and unique. Since these errors only differ on qubits five and six for the first stabilizer (i.e. the qubit support of the first stabilizer generator is only affected by qubits five and six), the following sequence gives fault-tolerant correction rules: we measure the same stabilizer generator again (in this case, $IIIXXXX$); then, we measure its conjugate ($IIIZZZZ$); finally, we measure the third stabilizer as follows. If the $IIIXXXX$ measurement (without flag) returned $(s=0)$ as the measurement outcome, we measure the stabilizer $ZIZIZIZ$; otherwise, we measure $XIXIXIX$. By analogy, our new sequences reduce the total stabilizer measurements from $6$ down to $3$ in the corresponding branches of the flag protocol (a reduction in the second subround by $50\%$). Note that the decoding step is in the conventional CSS style, which decodes $Z$ and $X$ errors separately. This protocol only utilizes the standard stabilizer generators.

\begin{figure}
\centering
\includegraphics[width=\columnwidth]{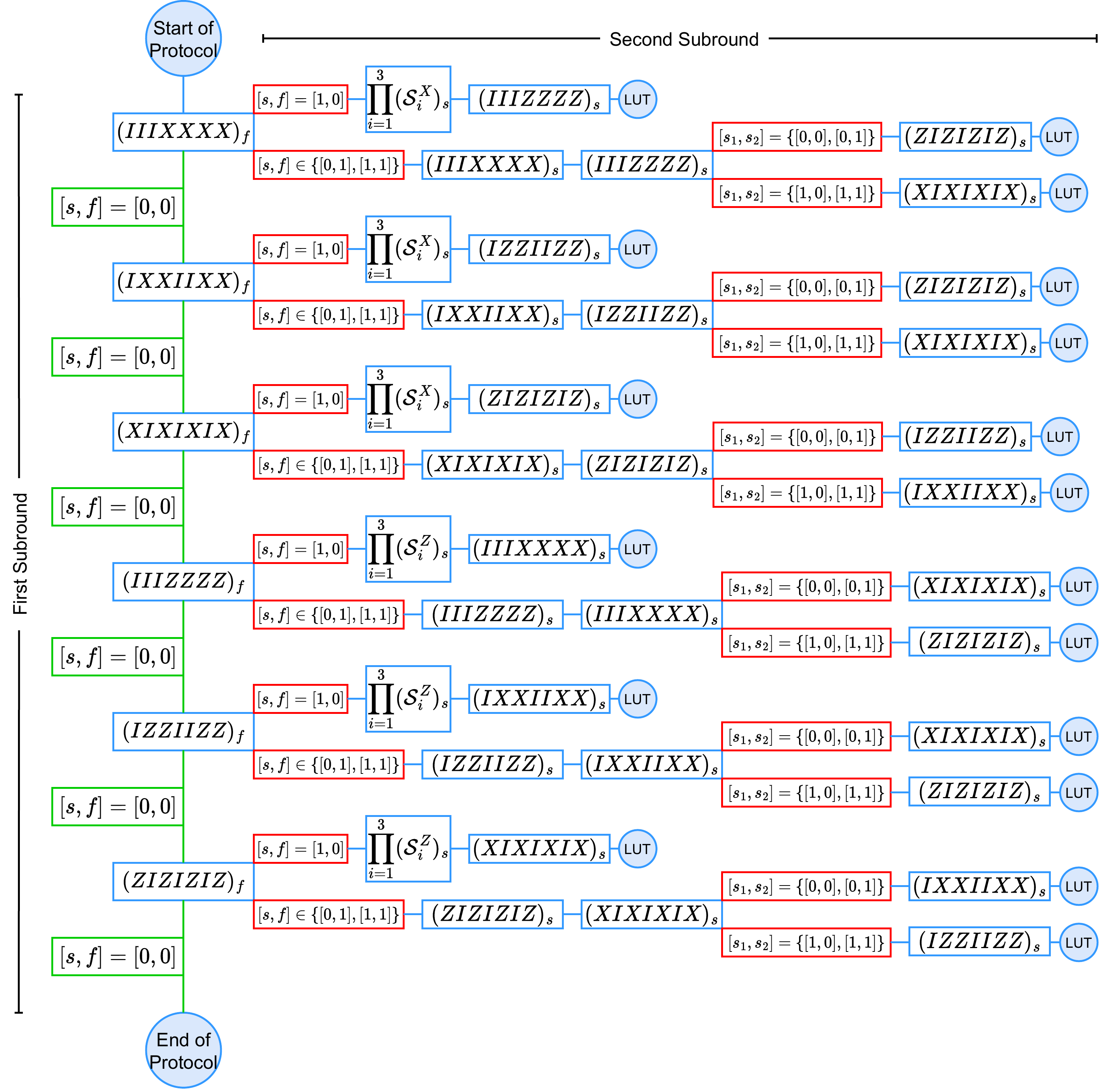}
\caption{The first new protocol introduced for the Steane code. Here, we modify the second subround of the flag protocol for both the cases of the flag firing or the syndrome firing. In the worst case, this reduction gives a lower gate count by $33\%$.}
\label{fig:newsteaneprotocol1}
\end{figure}

One can perform a further reduction in circuit depth for the cases when flag-qubit measurement outcomes are trivial, but the syndrome-qubit measurement outcome is non-trivial (i.e. the case of $[s,f] = [1,0]$). For instance, if the first stabilizer ($IIIXXXX$) measurement with the flag gadget returns a trivial flag-qubit measurement outcome but a non-trivial syndrome-qubit measurement outcome, then, the minimum set of errors which could have resulted in this measurement outcome are single-qubit Pauli errors $IIIZIII$, $IIIIZII$, $IIIIIZI$, $IIIIIIZ$, $IIIYIII$, $IIIIYII$, $IIIIIYI$, $IIIIIIY$, or a measurement error. To uniquely distinguish these nine possibilities in a manner ensuring fault tolerance, we modify the conventional sequence of six stabilizer measurements: all three $X$-type stabilizer generators ($IIIXXXX$, $IXXIIXX$, $XIXIXIX$) are measured to decode the $Z$ component of the errors unambiguously; next, only one more bit of information is required to distinguish between pure $Z$ or $Y$ errors, since the possible locations for these errors detected by the first stabilizer are the same. This information is provided by measuring a single (conjugate) stabilizer $IIIZZZZ$, which is trivial for the $Z$ errors, and non-trivial for the $Y$ errors. This logic extends to the other stabilizer measurements in a straightforward manner, and brings down the number of stabilizer measurements from $6$ to $4$ (a reduction of $33\%$). One should note that for the $[[5,1,3]]$ code, this line of reasoning regarding syndrome-qubit measurement outcomes would require measuring $4$ stabilizers, which would give no reduction compared to measuring the $4$ standard stabilizer generators; as such, we conclude that this approach may not be capable of reducing gate overhead further for the $[[5,1,3]]$ code. The Steane-code protocol is graphically depicted in \cref{fig:newsteaneprotocol1}.

\begin{figure}
\centering
\includegraphics[width=7cm]{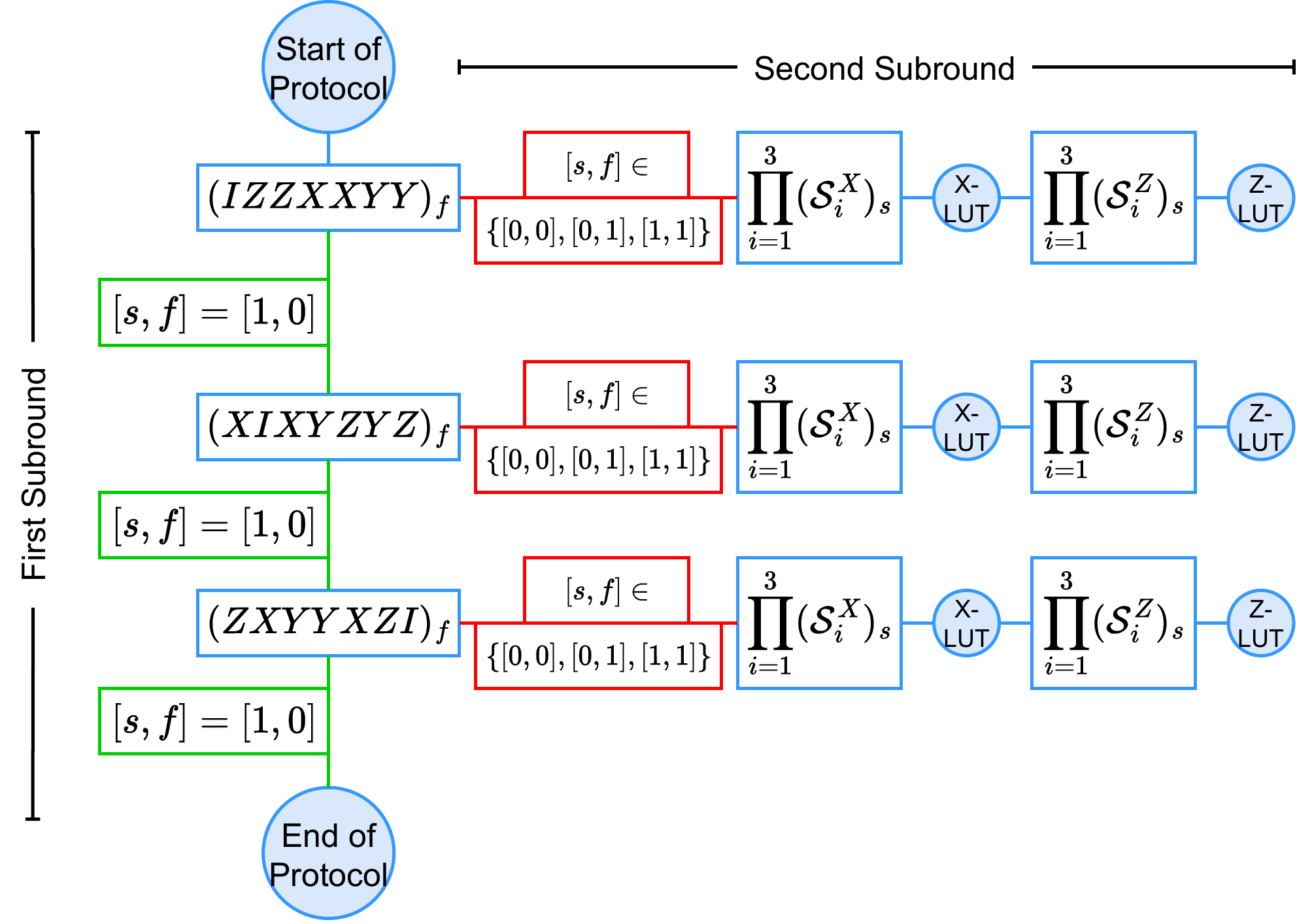}
\caption{The second new protocol introduced for the Steane code. Here, we modify the first subround of the flag protocol, instead of the second, by measuring three flagged $w=6$ stabilizers. This reduction lowers the two-qubit gate count by $33\%$ in the worst-case scenario, but is only lowered when compared with two or more $w=4$ generators measured in the first subround.}
\label{fig:newsteaneprotocol2}
\end{figure}

Finally, one may additionally modify the first subround, which contains flag qubits in our approach. Here we present an alternative FT protocol for the Steane code, which utilizes three $w=6$ stabilizers, instead of the typical six $w=4$ generators. This new protocol is shown in \cref{fig:newsteaneprotocol2}. In this way, we guarantee that a reduction is possible in the worst-case scenario, from six $w=4$ measurements with a total of $36$ gates, to three $w=6$ stabilizers with a total of $24$ gates (a reduction of $33\%$). Additionally, our first-subround approach reduces the number of measurement outcomes from $6$ to $3$, half as many as in \cite{chaoreichardt18}. Still, there are several cases where this protocol does not reduce gate overhead. For example, one may utilize less two-qubit gates in the case of only the first generator yielding non-trivial syndrome outcomes for the original protocol; in the first subround, our new protocol would utilize $8$ two-qubit gates, compared with $6$ for the original. On average though, one can expect high savings in terms of gate overhead, as all other first-subround cases lead to gate reductions.

\section{Results} \label{section:results}

\begin{figure}
\centering
\includegraphics[width=\columnwidth]{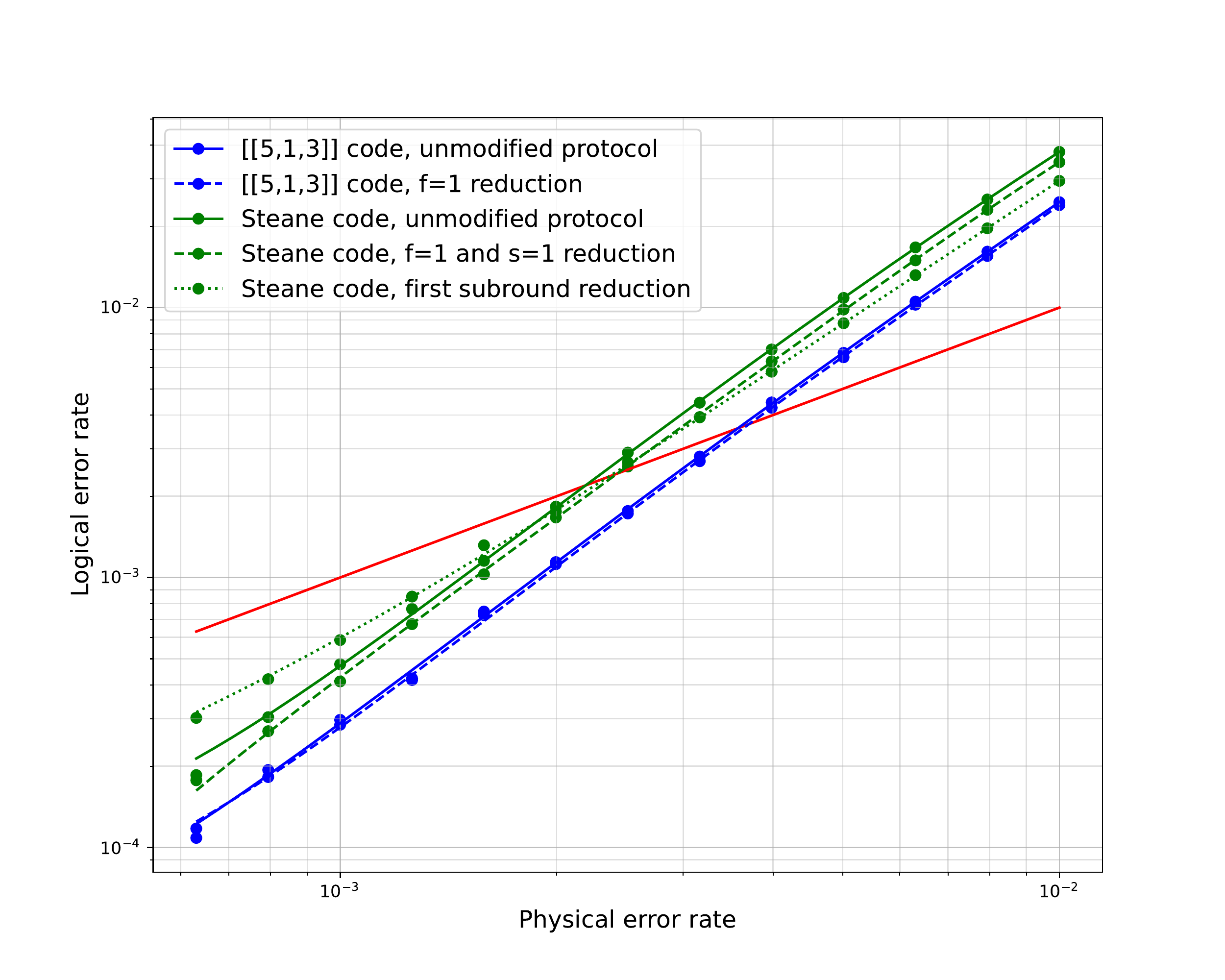}
\caption{Pseudothresholds results from the Qiskit simulation; the x-axis denotes the physical error rate $p_{\text{phys}}$, and the y-axis depicts logical error rate $p_{L}$. The pseudothresholds obtained in our work are better than the protocol used in \cite{chaoreichardt18}. However, our protocols use significantly less two-qubit gates. The pseudothreshold values found were: $3.5729 \times 10^{-3}$ for the unmodified $[[5,1,3]]$ code (\cref{fig:[513]_decisiontree_CR18}); $3.7030 \times 10^{-3}$ for [[5,1,3]] code with the $f=1$ reduction, an improvement of $3.64\%$ (\cref{fig:[513]_decision_new}); $2.1927 \times 10^{-3}$ for the unmodified Steane-code protocol (\cref{fig:steane_decisiontree_CR18}); $2.4302 \times 10^{-3}$ for the Steane code with the $f=1$ and $s=1$ reductions, an improvement of $11.16\%$ (\cref{fig:newsteaneprotocol1}); and $2.3611 \times 10^{-3}$ for the Steane code with the first-subround reduction, an improvement of 7.68\% (\cref{fig:newsteaneprotocol2}). For these simulations, error bars were too small to be seen easily in the plot. These simulations are freely available and open-source via \href{https://github.com/dhruvbhq/lowdepthflagqec}{https://github.com/dhruvbhq/lowdepthflagqec}.}
\label{fig:qiskit_pseudo}
\end{figure}

Our simulations were performed on the DelftBlue supercomputer \cite{delftblue}. More explicitly, we used: 384 cores from Intel Xeon compute nodes, each of 3.0 GHz; 192 GB of memory; and 480 GB of hard-drive space. All protocols were performed exactly as described, except for an extra perfect round of stabilizer measurement and correction, as in \cite{chaoreichardt18,arbitrarychamberland,gd16extraround}. For our simulations, we utilized a Monte Carlo wavevector simulation using Qiskit \cite{Qiskit} with $10^{7}$ trials per data point above physical error rate $p_{\text{phys}} = 10^{-3}$, and $10^{6}$ trials for $p_{\text{phys}} < 10^{-3}$, in order to save processor usage time.

Our noise model consists of independent and identically-distributed errors (iid). More specifically, we consider a circuit-level depolarizing noise model with the following parameters, in line with \cite{chaoreichardt18}: 

\begin{itemize}
\item With probability $p$, each two-qubit gate is followed by a two-qubit Pauli error drawn uniformly from $\{I,X,Y,Z\}^{\otimes 2} \backslash \{I \otimes I\}$.
\item With probability $\frac{4p}{15}$, preparation of the $\ket{0}$ state is replaced with $\ket{1} = X\ket{0}$. Likewise, a preparation of the $\ket{+}$ state would be equivalently replaced with $\ket{-} = Z\ket{+}$.
\item With probability $\frac{4p}{15}$, measurement outcomes are flipped. 
\item No idling error is considered.
\item The initial state encoding is considered to be noiseless. 
\item Single-qubit gates are taken to be noiseless.
\end{itemize}

Other more-sophisticated error models exist and have been tested with the flag protocol, in addition to several experimental realizations \cite{experiment1,experiment2,experiment3,experiment4}. We chose the error-model above since our goal is to compare against the original flag-protocol proposed in \cite{chaoreichardt18}. 

As shown in \cref{fig:qiskit_pseudo}, our proposals outperform those mentioned in \cite{chaoreichardt18}, particularly in the Qiskit simulations, where improvements of $3.64\%, 11.16\%,$ and $7.68\%$ were found for the new $[[5,1,3]]$ and both new Steane code protocols, respectively. Such an improvement comes of course as no surprise, as our protocols permit less opportunities for errors to propagate throughout the circuit. One may also consider error models which include idling noise \cite{arbitrarychamberland}, but we will save such exploration for future work. 

\section{Conclusion} \label{section:conclude}

In this work, we have investigated flag-style error correction and constructions of fault-tolerant syndrome-extraction circuit sequences, showing that elements from the full stabilizer group can permit reductions in overall gate overhead. We have focused on the $[[5,1,3]]$ and $[[7,1,3]]$ Steane codes, finding that reductions exist. As such, we report  lower two-qubit gate counts (per subround) by $25\%$ for the second subround's $f=1$ branch in the new $[[5,1,3]]$ protocol (\cref{fig:[513]_decision_new}), $33-50\%$ for the Steane code's $s=1,f=1$ protocol (\cref{fig:newsteaneprotocol1}), and $33\%$ for the Steane code's first-subround-reduction protocol (\cref{fig:newsteaneprotocol2}). In our pseudothreshold simulation, all reductions lead to concrete improvements over results from previously-established protocols. 

Our work also shows that much optimization is left at the level of quantum compilation before running stabilizer circuits on hardware; indeed, several studies have shown that topological-graph properties associated with circuit-level qubit interactions play an important role in suppressing errors for NISQ-era algorithms \cite{topologicalgraph,medina1,medina2,medina3}, and error-correction algorithms are no exception. It would be interesting to evaluate whether such lower-depth syndrome-extraction sequences as ours would be amenable still to NISQ-era devices, given their restricted connectivity, and whether or not current strategies such as parallel syndrome extraction \cite{gd3} could be utilized for further gate reductions.

Finally, recent research has proposed a unifying framework for fault tolerance, utilizing the ZX-calculus \cite{bombin2023unifyingZX}; this opens up the possibility for systematic optimization of stabilizer-measurement sequences. Answering this question will be the subject of future work.

\section{Acknowledgements}

We thank Aritra Sarkar for insightful discussions. MS and SF are grateful for support from the Intel Corporation. CGA is supported by the QuantERA grant EQUIP, by the Ministerio de Ciencia e Innovaci\'{o}n and Agencia Estatal de Investigaci\'{o}n, MCIN/AEI/10.13039/501100011033 and by the European Union “NextGenerationEU”/PRTR”. DE is supported in part by the JST Moonshot R\&D program under Grant JPMJMS2061.

\clearpage
\bibliographystyle{IEEEtran}
\bibliography{bibliography}

\end{document}